\documentclass[twocolumn,english,prl, showpacs, superscriptaddress]{revtex4-1}
\usepackage{color}
\usepackage{amsmath}
\usepackage{amssymb}
\usepackage{graphicx}

\makeatletter

\providecommand{\tabularnewline}{\\}
\usepackage{hyperref}
\hypersetup{
    colorlinks=true,
    linkcolor=red,
    citecolor=blue,
    filecolor=blue,      
    urlcolor=blue,
}
\makeatother

\usepackage{babel}
\begin{document}
\title{\noindent \textcolor{black}{A comment on a possible inadequacy of
new redefinitions of heat and work in quantum thermodynamics}}
\author{Eduardo B. Bottosso}
\address{Instituto de Física, Universidade Federal de Goiás, 74.001-970, Goiânia
- GO, Brazil}
\author{Jose S. Sales}
\address{Campus Central, Universidade Estadual de Goiás, 75132-903, Anápolis,
Goiás, Brazil}
\author{Norton G. de Almeida}
\address{Instituto de Física, Universidade Federal de Goiás, 74.001-970, Goiânia
- GO, Brazil}
\begin{abstract}
We analyze the new redefinitions of heat Q and work W recently presented
in {[}arXiv: 1912.01939; arXiv:1912.01983v5{]} in the quantum thermodynamics
domain. According to these redefinitions, heat must be associated
with the variation of entropy, while work must be associated with
variation of state vectors. Analyzing the behavior of two specific
examples, we show some peculiarities of these new redefinitions which,
based on the counterexample presented, seems to point to a possible
inadequacy of these redefinitions
\end{abstract}
\maketitle

\section{Introduction}

The pioneering work of Alicki \citep{Alicki1979}, who introduced
the concepts of heat Q and work W for a system interacting weakly
with a Markovian reservoir, strongly boosted the area of quantum thermodynamics,
not only with regard to its foundations but also concerning the study
of thermal machines \citep{Kosloff2013,Kosloff2017,Singh2020} and
their definitive limits. For example, efficiency and performance coefficient
have been extensively studied in different types of reservoirs \citep{RoBnagel2014,Long2015,Manzano2016,Klaers2017,Mendonca2020,Assis2021}.
The concepts of heat and work introduced by Alicki, however, needed
to be revisited and expanded to include situations not foreseen in
classical thermodynamics, as for example, initially correlated systems
\citep{Bera2017,Sapienza2019} and generalized reservoirs \citep{Bera2017,Manzano2018,Assis2019,Mendonca2020}. 

Although many of these formalisms have a sound theoretical appeal,
as they are recent, the case study is important for verifying their
compatibility with the framework of quantum thermodynamics as well
as if they reproduce results in accordance with quantum thermodynamics.
Recently, new redefinitions of heat and work, which aim to generalize
the redefinitions as introduced by Alicki, were independently proposed
by two groups \citep{Alipour2019,Ahmadi2020}. These redefinitions
are based on the following pillars: (i) the changes that occur in
a given system are taken into account in the corresponding reduced
density operator. (ii) Changes arising only in the state vector are
attributed to work, while changes arising from the entropy of the
system are attributed to heat. (iii) To calculate heat the von Neumman
entropy is to be used. In this work, we do a case study of these new
redefinitions and show some peculiarities that may possibly be indicating
an inadequacy of the formalism in \citep{Alipour2019,Ahmadi2020}. 

\section{\label{sec:II}New redefinitions for Work and Heat}

According to the formalism introduced by the authors in Refs. \citep{Alipour2019,Ahmadi2020},
the following protocol must be applied to a system interacting with
its surroundings, which we will generally call by reservoir: (i) Given
the initial states and the Hamiltonian, we evolve the composite system
to find $\rho_{AB}(t)$. (ii) From $\rho_{AB}(t)$, we diagonalize
$\rho_{A}(t)=tr_{B}\rho_{AB}(t)$ to obtain the eigenvalues $p_{i}(t)$
and eigenvectors $\left|\psi_{i}(t)\right\rangle $ in accordance
with $\rho_{A}(t)=\sum_{i}p_{i}(t)\left|\psi_{i}(t)\right\rangle \left\langle \psi_{i}(t)\right|$.
(iii) From the eigenvalues $p_{i}(t)$ and eigenvectors $\left|\psi_{i}(t)\right\rangle $
we calculate the work $W$ and heat $Q$ as given by $dQ_{A}(t)=tr\sum_{i}dp_{i}(t)\left|\psi_{i}(t)\right\rangle \left\langle \psi_{i}(t)\right|H(t),$
and $dW_{A}(t)=tr\sum_{i}p_{i}(t)d\left[\left|\psi_{i}(t)\right\rangle \left\langle \psi_{i}(t)\right|H(t)\right]$.

In this work we will focus on two simple models: (i) two qubits interacting
off-resonance with each other and (ii) a qubit interacting weakly
with its surrounding, as dictated by the standard master equation.
As we shall show, these two simple examples display a peculiar behavior
apparently not consistent with one would expect from a thermodynamics
analysis of these interacting system. In the following section we
will explore these examples in details.

\section{\label{sec:III}Results}

\subsection{Two qubits interacting off-resonantly}

The corresponding Hamiltonian model to the system A and B, including
the off-resonant interaction, is
\begin{equation}
H=\frac{\hbar\omega_{0}}{2}\sigma_{z}^{A}+\frac{\hbar\omega_{0}}{2}\sigma_{z}^{B}+\hbar g\sigma_{z}^{A}\otimes\sigma_{z}^{B},
\end{equation}
where $\hbar$ is the reduced Planck constant, $\sigma_{z}^{A(B)}$
is the Pauli matrix to the qubit A (B), $\omega_{0}$ is the transition
frequency between the two-level system, and $g$ is the coupling constant
of the interaction Hamiltonian. Note that this dispersive interaction
do not change the internal energy of the qubits; instead, the interaction
energy is stored in correlations produced during the time evolution.

Let us consider the following initial states for systems A and B \citep{Alipour2019,Ahmadi2020}:
\begin{equation}
\rho_{A}(0)=\begin{pmatrix}p & c\\
\bar{c} & 1-p
\end{pmatrix},\ \rho_{B}(0)=\begin{pmatrix}1/2 & 0\\
0 & 1/2
\end{pmatrix},\label{eq:initial condit}
\end{equation}
where in system A the parameter $c$ is only constrained by the positivity
of $\rho_{A}(0)$. 

Given these initial states, it is straightforward to obtain the reduced
density matrix to the evolved systems A:
\begin{equation}
\rho_{A}(t)=\begin{pmatrix}p & c\cos2gt\\
\bar{c}\cos2gt & 1-p
\end{pmatrix},\ \rho_{B}(t)=\begin{pmatrix}1/2 & 0\\
0 & 1/2
\end{pmatrix}=\rho_{B}(0),\label{eq:final}
\end{equation}

For our purpose, we diagonalize this system for $p=\frac{1}{2}$ and
$c=\bar{c}=\frac{1}{2}$. The eigenvalues and eigenvectors are found
to be
\begin{align}
\lambda_{+}(t)=p_{1}(t)=\sin^{2}gt\\
\lambda_{-}(t)=p_{2}(t)=\cos^{2}gt
\end{align}
\begin{align}
\left|\psi_{1}\right\rangle =\frac{1}{\sqrt{2}}\left[\left|0\right\rangle +\left|1\right\rangle \right] & ,
\end{align}
\begin{align}
\left|\psi_{2}\right\rangle =\frac{1}{\sqrt{2}}\left[\left|0\right\rangle -\left|1\right\rangle \right] & ,
\end{align}
and then, following the protocol described above we write $\rho_{A}(t)$
using its instantaneous eigenbasis:
\begin{equation}
\rho_{A}(t)=\sum_{i=1}^{2}p_{i}\left|\psi_{i}\right\rangle \left\langle \psi_{i}\right|=\sin^{2}gt\left|\psi_{1}\right\rangle \left\langle \psi_{1}\right|+\cos^{2}gt\left|\psi_{2}\right\rangle \left\langle \psi_{2}\right|.
\end{equation}
Note that, to these parameters, different from the eigenvalues, the
eigenkets do not depend on time. Also, note that $\sigma_{A}^{Z}\left|\psi_{1(2)}\right\rangle =\left|\psi_{2(1)}\right\rangle $.
Next step, we calculate both $W$ and $Q$ according to Ref. \citep{Alipour2019,Ahmadi2020}.
To the work $W$:

\begin{align}
dW_{A}(t)=\sum_{i}p_{i}(t)d\left[\left\langle \psi_{i}\right|H_{A}\left|\psi_{i}\right\rangle \right]=0 & ,
\end{align}
which is zero both because the eigenvector is time-independent and
because $\left\langle \psi_{i}\right|H_{A}\left|\psi_{i}\right\rangle =0$.
Therefore, as $W_{A}(t=0)=W_{A}(t=T)$, $W_{A}$ is zero for all t's.
To the heat $Q$:

\begin{equation}
d\left\langle Q_{A}\right\rangle =\sum_{i=1}^{2}dp_{i}(t)\left\langle \psi_{i}\right|H_{A}\left|\psi_{i}\right\rangle =0,
\end{equation}
because $\left\langle \psi_{i}\right|H_{A}\left|\psi_{i}\right\rangle =0$.
Therefore, as $Q_{A}(t=0)=Q_{A}(t=T)$, $Q_{A}$ is zero for all $t$'s.
Since the internal energy of each qubit is not varying, according
to the first law $\triangle U=\triangle Q+\triangle W=0$, we have
the peculiar situation where two systems are interacting while neither
heat nor work is being exchanged, which is precisely one of the examples
used to criticize the Alick's approach in Ref. \citep{Ahmadi2020}.
It is important to note that this result ($\triangle U=\triangle Q=\triangle W$)
is only true for the particular parameters we have used above; had
one choose $p\neq1/2$ , although $\triangle U=0$, one could find
$\triangle Q=-\triangle W\neq0$. Yet, from this example, it is to
be noted that since $p_{i}(t)'s$ vary, entropy is also varying. In
fact, from differentiating von Neumman entropy to system A we find

\begin{equation}
dS_{A}=-\sum_{i}^{d}dp_{i}(t)\ln p_{i}(t),\label{eq:Entropy1}
\end{equation}
or, after replacing $p_{1}(t)=\sin^{2}gt$, $p_{2}(t)=\cos^{2}gt$:

\begin{equation}
\frac{dS_{A}}{dt}=-\sin\left(2gt\right)\ln\left(\tan^{2}gt\right),
\end{equation}
and, after integration:
\begin{equation}
S_{A}(t)=-2\ln\left[\cos(gt)\right]-\sin^{2}(gt)\ln\left[\sin^{2}(gt)\right]/\cos^{2}(gt).\label{eq:entropy}
\end{equation}
This is another peculiar finds of this formalism: according to \citep{Alipour2019,Ahmadi2020},
entropy variation is connected with exchange of heat; however, we
have again this peculiar behavior where entropy varies without any
heat being exchanged with the surroundings, thus pointing to a possible
inadequacy of the formalism.

Let us now turn our attention to system B. Since the evolved state
remains unchanged for all times, this means that, different from system
A, system B do not exchange neither work nor heat during its evolution.
In other words: during systems A and B interaction there will be $dQ_{A}=-dW_{A}\neq0$
flowing out/into the system A in general. The only explanation left
is that the energy stored in the correlations flows only out/into
the system A alone, and system B works as a catalyst in this process.
Now, if we take the parameters to be $p=c=1/2$, we have this peculiar
situation: $dU_{B}=dQ_{B}=dW_{B}=0$, $dU_{A}=dQ_{A}=dW_{A}=0$, at
the same time that $S_{A}(t)\neq0$, i.e., there is no energy flowing
to any of the systems and yet the entropy of system A varies, which
should lead to a no-null heat according to statement (ii) in Introduction.

$\,$

\subsection{A single qubit dissipating}

The following master equation describes a qubit under a weak coupling
with its reservoir in the interaction picture:
\begin{multline}
\dot{\rho}=-\frac{\gamma}{2}\overline{n}\left[\sigma\sigma^{+}\rho-2\sigma^{+}\rho\sigma+\rho\sigma\sigma^{+}\right]\\
-\frac{\gamma}{2}\left(\overline{n}+1\right)\left[\sigma^{+}\sigma\rho-2\sigma\rho\sigma^{+}+\rho\sigma^{+}\sigma\right],\label{eq:ME}
\end{multline}
where $\sigma$ and $\sigma^{+}$ are the lowering and raising operators
to the system, $\overline{n}$ is the reservoir average thermal excitation
and $\gamma$ is the dissipation rate. This equation can be exactly
solved, resulting
\begin{equation}
\rho_{A}(t)=\begin{pmatrix}\rho_{ee}(t) & \rho_{eg}(t)\\
\rho_{eg}^{*}(t) & 1-\rho_{ee}(t)
\end{pmatrix},\label{qubit}
\end{equation}
where 

\begin{equation}
\rho_{ee}(t)=\frac{\left[\left(2\overline{n}+1\right)\rho_{ee}(0)-\overline{n}\right]e^{-\left(2\overline{n}+1\right)\gamma t}+\overline{n}}{\left(2\overline{n}+1\right)},
\end{equation}
and

\begin{equation}
\rho_{eg}(t)=\rho_{eg}(0)e^{-\left(\overline{n}+\frac{1}{2}\right)\gamma t}.
\end{equation}

\begin{figure}
\centering{}%
\begin{tabular}{cc}
\includegraphics[width=6cm,height=6cm]{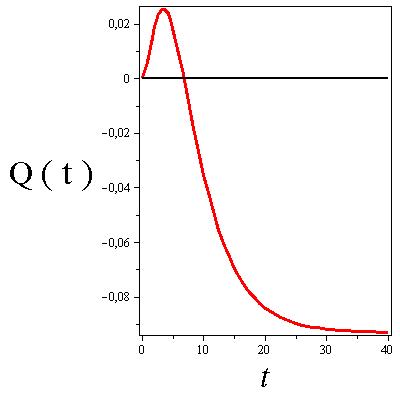} & \tabularnewline
\end{tabular}\caption{\label{Fig heat} The heat exchanged by a qubit in a superposition
state interacting with a heat bath at zero absolute temperature. Note
that initially the qubit apparently absorbs energy from the vacuum
reservoir.}
\end{figure}

Next, we diagonalize Eq.(\ref{qubit}) and find heat according to
$d\left\langle Q\right\rangle =\sum_{i=1}^{2}dp_{i}(t)\left\langle \psi_{i}\right|H\left|\psi_{i}\right\rangle $,
where $H=\frac{\hbar\omega_{0}}{2}\sigma_{z}$, and choose the qubit
initial condition as being the following pure state:
\begin{equation}
\rho(0)=\frac{1}{2}\begin{pmatrix}1 & 1\\
1 & 1
\end{pmatrix}.
\end{equation}
In Fig. (\ref{Fig heat}) we plot $Q(t)$ \emph{versus} t for $T=0$.
Note the following peculiar behavior: the heat of the system is greater
than zero, thus the system is, for a while, drawing energy from the
vacuum. One could argue that since according to the first law $\triangle U=\triangle Q+\triangle W$,
and since $\triangle U<0$ always, the net flux is from the system
toward the reservoir, thus indicating that energy flows from the system
to the vacuum reservoir. However, if we pay attention in what a true
reservoir should be, i.e., an infinite collection of quantum harmonic
oscillators as modeled by Eq.(\ref{eq:ME}), the recurrence, which
is the time to the energy goes back to the reservoir, should be infinite,
and no energy should go from the vacuum to the system \citep{RoBnagel2014,Long2015,Manzano2016,Klaers2017,Mendonca2020,Assis2021}.

\section{\label{sec:V}Conclusion}

In this work we did two case studies in which we applied the formalism
that introduces new definitions of heat and work in quantum thermodynamics
as introduced in Refs. \citep{Alipour2019,Ahmadi2020}. One of the
examples studied consists of two interacting qubits, a situation in
which the concepts of heat and work are difficult to generalize, and
there is currently no consensus on which formalism correctly describes
the corresponding thermodynamic behavior, in particular the heat and
work exchanged by the systems. The second example consists of qubit
interacting weakly with a thermal reservoir in Born-Markov approximations,
as described by the usual master equation, and where it is well known
that the formalism developed by Alick correctly describes heat and
work. The analysis of the first example shows that, for certain initial
conditions, the entropy of a system can vary without a counterpart
in the heat variation, which is in apparent contradiction with one
of the assertions on which the formalism \citep{Alipour2019,Ahmadi2020}
is based. In the second example, the formalism \citep{Alipour2019,Ahmadi2020}
shows a heat variation in the qubit that is apparently incompatible
with what is expected of a genuine reservoir at zero absolute temperature,
which must always absorb heat from the system in which it is in contact.
These two examples seem to point to a possible inadequacy of the formalism
in \citep{Alipour2019,Ahmadi2020}.

\bibliographystyle{apsrev4-1}
\bibliography{References}

\section*{Acknowledgments }

We acknowledge financial support from the Brazilian agency, CAPES
(Financial code 001) CNPq and FAPEG. This work was performed as part
of the Brazilian National Institute of Science and Technology (INCT)
for Quantum Information\textbf{ }Grant No. 465469/2014-0.
\end{document}